\documentclass{neuthist18}

\def\be{\begin{equation}}
\def\ee{\end{equation}}
\def\bea{\begin{eqnarray}}
\def\eea{\end{eqnarray}}

\newcommand{\Photo}{}

\begin{document}
\vspace*{4cm}
\title{Neutrino Physics in Historical Context}

\author{Q. Rodriguez}

\address{Universit\'{e} Clermont Auvergne, PHIER, F-63000 Clermont-Ferrand, France\\Maison des sciences de l'Homme, PHIER, 4 rue Ledru, 63057 Clermont-Ferrand cedex 1, France}

\maketitle\abstracts{
This contribution aims to give an overview of the historical context of neutrino physics. I will present the strong social trends that shaped physics and the way physicists worked, along the 20th century. First, we will see the background of the birth of nuclear physics in the interwar period. Then, we will examine the deep implications the Second World War had, to conclude with the specificities of post-war years for nuclear and particle physics.}

\section{Early Neutrino Physics: The Interwar Period}

I have to say, to introduce this short paper, that I am not a specialist in particle physics or particle physics history. So, what could a generalist historian of science bring to a conference full of experts in neutrino physics? Maybe a way to shift the focus a bit. I will consider it as my job here to convince you that the history of the neutrino, is not only a history of great intellectual discoveries and wonderful new techniques, but also from end to end a history of the 20th century. A history of nations, politics and war. A history of institutions and social changes, all deeply intertwined, that can enlighten the strong trends inside which neutrino physics has evolved, and will continue to.

This history starts during the interwar period, as several authors have reminded us. It is precisely at this time that the way of doing science, and physics especially, started to resemble something we know nowadays. Actually, the interwar period was a period of major changes for science that we cannot understand without calling back to the beginning of the century and particularly the First World War.

Let us remind ourselves briefly what physics looked like at the turn of the century. An overwhelming majority of physicists were working in Western Europe. American universities were still small and distant institutions, in a quickly developing but isolated country. Most scholars, who were not yet named \emph{researchers}, were still amateurs, living off of personal resources, teaching or business activities, or patronage and various grants. Work was essentially individual or organized between small groups of people, and what we could call ``laboratory work'' was often led at private residences and not in universities. Leading papers were written in German or
French\footnote{Even later, early 1934, Enrico Fermi wrote his famous paper on beta-decay theory simultaneously in Italian (\textit{Nuovo Cimento}) and in German (\textit{Zeitschrift f\"ur Physik}). It has been translated in English much later.}. And science was mainly seen among society as an enjoyable inquiry with moral involvement more than a material or economic one.

Finally, the existence of atoms had just been completely accepted, and the physics of radiation was bursting into an explosion of new observations the likes of which fundamental physics had not known for a while. The experiments stemming from this whole new continent became the building blocks of quantum theory.

In this landscape, the First World War was probably the event having the most profound consequences on 20th-century science, on par with or maybe even having more impact than the Second World War.

Scientists played a very important role throughout the war. Aviation was used for the first time in a conflict. Submarine use became significant. Chemists developed poisonous gases. Heavy artillery demanded more and more ballistic computations. And meanwhile, the twice nobelized Marie Curie developed medical radiology to help the Red Cross treating the countless injured soldiers. At the same time, the total war, produced by the clash of the imperialisms among European states, involved the entire societies, resting upon nationalist ideologies. All of this led ruling elites and each nation's society to think of their scientists as a highly useful tool for power. And even if we usually date the end of the First World War in 1918, this schema actually persisted more or less until the Second World War.

What have been the consequences for physics? Firstly, the professionalization of scientists. The importance of science for national power, and the new prominence taken by the state in each nation's society together made science a political and national aim. At the same time, the terrible social conditions of the war aftermath had broken the fragile equilibrium between personal resources and private fundings that once had permitted some scientists to live and work. The time had come to construct national public scientific organizations with staff researchers. In France, it was the CNRS, \emph{Centre national de la recherche scientifique}, planned by the government of the Popular Front and the first undersecretary for scientific research, Ir\`{e}ne Joliot-Curie, daughter of Marie Curie and recipient of the 1935 Nobel Prize in Chemistry for the discovery of artificial radioactivity. So, scientific research became a professional career during this interwar period, something it had never really been before in European history.

Secondly, another direct outcome of this new social place for science ---and especially physics--- was its incredible penetration into society as a whole, hand in hand with the rise of advertisement (or propaganda) and the birth of consumerism. Modern physics became synonymous with the power of a nation and with a radiant future. Institutions dedicated to popularization of science were built, like the \emph{Palais de la D\'{e}couverte} designed by Jean Perrin (Nobel Prize in Physics in 1926) in Paris. In France, the discovery of radium by Marie Curie inspired in just a few years a now absurd trend of radium \emph{consumption}. Advertisements were praising radioactive water, or face cream; very good for the complexion, guaranteed! Or even radium talc for babies!

Finally, the war had been the accelerator that propelled the American economy to a worldwide leading position. The center of gravity for physics was still in Germany, but the rapid development of scientific institutions in the United States made the physics community experience a first kind of globalization across the Atlantic Ocean. The most famous example of this internationalization of physics was the organization in Brussels, by the Belgian industrial Ernest Solvay of a series of conferences, known as the ``Solvay conferences'', more or less every three years starting from 1911. A large part of the animated debates about the nature of the quantum world and the correct theories to understand it took place there. In a way, these meetings were the first international conferences of theoretical physics. Except that at the time, basically, all theoretical physicists in the world could fit in a single room.

\section{Nuclear Physics and the Second World War}

We saw how deeply the way physicists' work changed during the interwar period, and how these changes were driven by a new role assigned to science by society. We will now address the next main topic. The topic of the Second World War, obviously. This Second World War is so much a pursuit of the first one, that it will of course point mostly towards the same directions.

One of the most impressive consequences was the tipping of the center of gravity for physics, from Germanic countries to the United States, over the course of ten years. The events we saw in the previous part had already laid out the field for the welcoming of a major part of international physics on the other side of the Atlantic. But the trigger was obviously the rise to power of the Nazis in Germany in 1933. Just among Jewish physicists, physicists with Jewish family, or left-wing political opponents, the Germanic countries saw in a few years a whole part of their scientific community fleeing central Europe. Erwin Schr\"{o}dinger chose Ireland, Max Born the United Kingdom and Lise Meitner Sweden. But for most of them, the United States had become the most natural destination. Among the most famous ones, we could name Albert Einstein, Niels Bohr, Enrico Fermi, Wolfgang Pauli, Hans Bethe and Leo Szilard. And most of them \emph{stayed} in the United States once the war was over. Europe was then a devastated battleground, particularly Germany, who was placed under Allied administration. And more decisive, the economic domination of the United States became far stronger than it had become in the wake of the First War. For the second time in thirty years, the United States became the creditor of the European war effort, and decided to fund the reconstruction in exchange for massive import of American products. This was known as the Marshall Plan.

At the same time, the combination of the national scientific institutions of the interwar period with the abrupt conversion of countries to war economy, especially in the United Kingdom and the United States, led to an incredible intensification of the previous trends. Historians of science have called this new regime of science, completed with the Second World War, \emph{Big Science}. How Big? Big in several meanings. First, because of a quantitative explosion. Of the number of scientists as well as the national spendings. Estimated research spendings for the United States were multiplied a thousandfold between 1935 and 1945. Second, because of the organization of scientific work in big laboratories mobilizing hundreds to thousands of researchers. And finally, because of the organization of this work around large instruments, like particle accelerators, large telescopes or computers.

Since the states were generously funding these activities, it was decided to fix practical goals to scientists, decided by a central administration, and to organize research in collaborative units, not unlike industrial organization. Around these practical goals that could help to win the war, multidisciplinary teams were established, with almost unlimited funds. All across the United States and the United Kingdom, scientists were \emph{enlisted} like this, with the idea that scientific achievements could decide the outcome of the world conflict.

And in a way it did. The leading position acquired by Great Britain in the development of RADAR technologies during the conflict were of the highest importance for the Navy and the Air force. The field of operations research appeared, combining computation and mathematical analysis, to help solve the huge organizational issues of the war. Cybernetics and information theory arose from the study of communication problems and automatic guidance for air defense systems. And we have to mention the development of digital computers, for ballistic calculations or codebreaking. The movie \emph{The Imitation Game} about the work of Alan M. Turing during the war recently popularized the importance of cracking German intelligence codes for defeating the German naval blockade of Great Britain.

And, above all, ``the bomb'', obviously. I will not stay long on this well-known story. Just a few remarks on the importance of this episode for fundamental physics development. From the first publication on nuclear fission in 1939, to the Hiroshima bombing, only six years had passed. During this interval, the handful of nuclear physicists in the world informed of these works alerted their governments about the potential military use of this fundamental discovery. All open publication on this topic stopped. And in 1945, the most frightful weapon ever built by humankind killed 70,000 people instantly, and 200,000 more during the next five years, in one unique explosion. I think there is no need to stress more the importance took by fundamental physics in the 20th century.

The team of Fr\'{e}d\'{e}ric Joliot-Curie, in Paris, was probably the most advanced on this subject in 1939, but he preferred to stay in the occupied France and all their work stopped until the end of the war~\cite{latour}. Meanwhile, Robert Oppenheimer was charged by Roosevelt to be the scientific leader of the Manhattan Project. A large number of the most brilliant physicists, who had fled central Europe just some years ago, worked with him on the atomic bomb, with the assistance of more than \emph{one hundred and fifty thousand} engineers, technicians and workers across the United States. The physicists' team founded the laboratory of Los Alamos ---a kind of secret city built in the middle of nowhere in New Mexico--- which basically became the prototype of modern fundamental physics laboratories. For the study of the nuclear detonation, it is there that the mathematician Stanislaw Ulam invented the Monte Carlo methods and proceeded to the first computer simulations in history.

For scientists, this Big Science meant a loss of their autonomy. They were no longer the ones to decide on their research priorities, instrument choices or colleagues. Concurrently, military power was not, for the first time, a demographic issue, but a scientific one. And if the state could discipline and organize scientific research, this scientific issue could become itself an economic one. This situation led to a kind of historical mutual compromise between scientists and political power. This compromise was set out in the most famous and clearest way by Vannevar Bush, MIT researcher on computers and scientific advisor of then President Roosevelt, in a public report to the President of the United States published in 1945. Its title was \emph{Science: The Endless Frontier}~\cite{bush}.

Bush was requesting the perpetuation of the national budgets without limit, like those they had known during the War, but argued for an ideal of autonomy for ``pure science''. In exchange, society and political power could count on rapid return on investment, in terms of economic growth and military applications. With the example of nuclear physics in mind, fundamental physicists encountered no difficulty to convince military authorities to pay huge amounts to let them investigate whatever research lines they would like. Something big and useful would certainly come out of this one day or another!

Obviously this solution established itself thanks to the context of the Cold War, as we are going to see in the last part. Indeed, we have not talked at all about the USSR until now, which however was maybe the first country to enter in this Big Science regime, despite its economic weakness compared to the United States.

\section{The Cold War, Golden Age of Particle Physics?}

We saw the compromise between scientists and political power at the close of the War that historians call Big Science. This compromise had been stabilized by the constant preparedness for war that the United States, Europe, and the USSR continued to experience during thirty years. This period is obviously what we know as the cold war, and was also a kind of Golden Age for nuclear and particle physics, alongside space sciences, that ensured spendings to continue to flow and the number of scientists involved to grow. Spendings were coming from the states, and from national companies, since these countries knew during this period a constant economic growth borne by the states protection and investments towards national industries. Eisenhower dubbed this intricacy among the state, scientific research, industry, and military goals in a famous speech in 1961 as ``the military-industrial complex''.

This led to the invention of the concept of \emph{research \& development} in these big national companies. The idea that we could apply the methods of organization of scientific research invented during the war to innovation for the industry, with cycles from fundamental research to industrial applications. In the United States the most famous example was Bell Labs, created by the monopolistic phone company AT\&T. From Bell Labs came an impressive number of Nobel Prizes and breakthroughs that contributed in a major way to communication technologies and material sciences. We could practically say that the modern field of condensed matter physics emerged there. To give an idea, we could name Claude Shannon, who wrote there his mathematical theory of communication during the war, Arthur Schawlow and Charles Townes who built the first Laser in 1960 (in parallel with a Soviet team), Arno Penzias and Robert Wilson who discovered the cosmic microwave background in 1964, or John Bardeen William Shochkley and Walter House Brattain who produced in 1947 the invention that maybe changed the second half of the 20th century in the most profound way: the transistor, that gave birth to the whole field of microelectronics.

In France, the best example of this intricacy between industry, fundamental physics, and military goals, was the creation of the \emph{Commissariat \`{a} l'\'{e}nergie atomique} (CEA) in 1945, under the direction of the famous physicist Fr\'{e}d\'{e}ric Joliot-Curie and Raoul Dautry, former minister for weaponry. The first goal of the CEA was to take over the work Joliot-Curie and his team stopped in 1939 on nuclear fission, and be among the first after the Americans to produce a reactor based on controlled nuclear chain fission. This was done in 1948.

In Europe, the intricacy between political goals, industry, and science took a specific shape with the construction of a European union. Far from nowadays European Union, the aim was to create an intergovernmental space between the states that were opposed during the conflict, to share strategic technologies and resources, and thus ensure the end of the national rivalries that made the first half of the century. This was done with the European Coal and Steel Community, pooling these strategic industries for war economies together. And of course with nuclear science and industry with the EURATOM organization and the CERN.

I will not say too much about the CERN, because a lot of people contributing to this conference know far better than me this story. But I think I have set the scene to see more clearly the very special historical moment that allowed this unique project: the sharing of the most frightening science and technologies that came out of the war, between former enemies, for civil uses.

However, this compromise of the Big Science was a critical deal for a lot of scientists. Hiroshima had been, of course, a moral rupture for a lot of physicists. Fundamental science would no longer be considered as a neutral and purely intellectual adventure. Fundamental science \emph{can} be harmful by itself. On their side, medicine and biology had Auschwitz. Eugenics had been invented as a medical specialty, with its journals and learned societies, advocating for mass sterilization for the greater good. And the trial of the Nazi doctors who enjoyed to experiment freely on humans in the extermination camps led to the Nuremberg Code in 1947 ---the first international text for ethics principles restricting human subject experimentations.

In the same movement, nuclear weapons had become the \emph{physicists'} burden. To index the freedom of physicists to decide on their research lines, on the power capacity they could give to their country starts to be a problem, when the consequences of their research can be so big and tragic. And so, a lot of physicists, even those who worked on the Manhattan Project, started to join the pacifist movement and campaigned for denuclearization. Not all of them, of course. John von Neumann or Edward Teller fell on bad terms with their former colleagues for their support of the thermonuclear bomb program. But Oppenheimer and Sakharov ---the leader of the soviet nuclear program--- made their public soul-searching some years after. Leo Szilard, who wrote the famous letter to Roosevelt signed by Einstein in 1939 that led to the Manhattan Project, initiated a petition in July 1945, signed by a hundred of physicists, to ask president Truman to use the bomb against Japan only if surrender discussions would not succeed. This petition was banned, but Szilard and Einstein created the Emergency Committee of Atomic Scientists the year after to publicly contest the development of nuclear weapons. Linus Pauling won fame campaigning alongside Szilard and Einstein, and finally became recipient of the Nobel Peace Prize in 1962. Joliot-Curie, in France, initiated the international Stockholm Appeal in 1950, calling for an international ban on nuclear weapons. He was then dismissed from the head of the CEA, and after his departure, the CEA finally engaged in the production of the French atomic bomb.

I would like to mention one last important feature of this period, concerning especially particle physics. The quantitative expansion and the new way to organize big collaborations between different specialties, born in the war, led to an impressive diversification of subspecialties, particularly in microphysics. The historian of science Peter Galison produced a masterful analysis of this evolution~\cite{galison}. He showed how different communities and experimental traditions came to work together in huge collaborations around big instruments like the accelerators. These \emph{subcultures} can be organized around experimental, theoretical, instrument-making, or data analysis goals. Each of these communities follows its own scientific aims, and works in a way modeled by its own material culture, but develops communication interface with the others, that Galison dubbed \emph{trading zones}, where a specific language, mixing vocabularies coming from the different scientific subcultures, is used. This fragmentation-cooperation, very specific to contemporary physics, dissolved the very notion of the author of a scientific result, as a lot of the contributions to this conference show.

\section*{Acknowledgments}
This paper do not pretend to be original. It is essentially an attempt to give a big picture, based on many well-known works among historians of contemporary science. Readers will find in the references~\cite{pestre,krige} two valuable overviews of this material that I found useful to prepare this contribution. Finally, a special mention to Taya Flaherty who thoroughtly proofread the English writing of this paper.

\end{document}